# Observation of giant nonvolatile magneto-thermal switching in superconductor-ferromagnet hybrids

Yui Sakamoto[1,2], Fuyuki Ando[1*], Keigo Ito[3], Hossein Sepehri-Amin[1], Takamasa Hirai[1,3], Yuto Watanabe[4], Poonam Rani[4], Yoshikazu Mizuguchi[4], and Ken-ichi Uchida[1,2,3*]

[1]*Research Center for Magnetic and Spintronic Materials, National Institute for Materials Science, Tsukuba 305-0047, Japan*

[2]*Graduate School of Science and Technology, University of Tsukuba, Tsukuba 305-8573, Japan*

[3]*Department of Advanced Materials Science, Graduate School of Frontier Sciences, The University of Tokyo, Kashiwa 277-8561, Japan*

[4]*Department of Physics, Tokyo Metropolitan University, Hachioji 192-0397, Japan*

E-mail: ANDO.Fuyuki@nims.go.jp, UCHIDA.Kenichi@nims.go.jp

**Abstract**

Magneto-thermal switch is a crucial thermal component which enables heat transfer control by the application of an external magnetic field. Recently, a nonvolatile behavior in magneto-thermal conductivity at zero magnetic field was observed in type-II and phase-separated superconductors owing to magnetic flux pinning nature, leading to an energy-efficient thermal control technology. However, the nonvolatile magneto-thermal switching ratio has been much lower than the volatile one in conventional materials. Here, we demonstrate a giant nonvolatile magneto-thermal switching in ferromagnetic Fe-superconducting Pb hybrids. The dispersion of pure Fe particles realizes increased electron and decreased phonon contributions in the thermal conductivity, which enhances the magneto-thermal switching ratio at the superconducting-to-normal conducting phase transition. Furthermore, in concert with trapped magnetic flux by supercurrent, ferromagnetic moment of Fe breaks the superconductivity of Pb matrix at zero magnetic field, enabling a significantly large nonvolatility even with a slight amount of Fe inclusions. Consequently, the nonvolatile magneto-thermal switching ratio reaches 719% in maximum at the Fe ratio of 8.7 vol%, which is more than twice the previous record value observed in Pb-Sn composites and the volatile one in pure Pb. This work broadens the exploration space and strategy for giant nonvolatile magneto-thermal switching materials.

## 1. Introduction

Magneto-thermal switch (MTS) is a solid-state thermal component to control heat transfer between two ends by tuning its thermal conductivity $\kappa$ via the application of an external magnetic field.[1,2] Whereas classical thermal switches control heat transfer via physical contact or flow of gases and liquids, MTS operates without any moving parts, which ensures superior cycling durability. The fundamental mechanisms of MTS have been intensively studied and proposed involving the magneto-thermal resistance effects,[3–5] magnon thermal conduction,[6–10] and superconducting phase transition.[11–17] One of the important issues on MTS is to increase the magneto-thermal switching ratio (MTSR) to properly turn on and off heat transfer, which is defined as,

$$\mathrm{MTSR}(T,H) = \frac{\kappa(T,H) - \kappa(T,H = 0\ \mathrm{Oe})}{\kappa(T,H = 0\ \mathrm{Oe})}. \qquad (1)$$

Here, $T$ is absolute temperature and $H$ is an external magnetic field.

MTS based on metallic superconductors is promising for cryogenic temperature applications owing to their giant MSTR over 2000% below critical temperatures $T_c$.[13–15,17] In normal metals, $\kappa$ is the sum of electron and phonon contributions ($\kappa_{el} + \kappa_{ph}$). According to the Bardeen-Cooper-Schrieffer theory, when a metal transitions from a normal conducting state into a superconducting state, conduction electrons form Cooper pairs which do not contribute to thermal conduction, resulting in the disappearance of $\kappa_{el}$ and reduction of $\kappa$. Thus, $\kappa$ for metallic superconductors can be dynamically tuned by the application of $H$ across the critical field ($H_c$) associated with the transition between superconducting and normal conducting states. Owing to its large MTSR, metallic superconductor-based MTS has been practically used in adiabatic demagnetization refrigerators, for example, at NASA's Goddard Space Flight Center.[2]

Recently, a nonvolatile characteristic of MTS has been observed in type-II and phase-separated superconductors, which will lead to a new class of heat transfer technology with lower power consumption than that of conventional MTS. Both type-I and -II superconductors are known to show hysteresis in magneto-thermal conductivity when an intermediate-mixed state is formed around $H_c$.[11,12,18,19] Arima et al. and Yoshida et al. firstly reported the nonvolatile characteristic that the hysteresis sustains even at zero magnetic field in type-II superconductors Nb and V, respectively.[14,15] The nonvolatile MTSR is defined as,

$$\text{Nonvolatile MTSR}(T,H) = \frac{\kappa_{\mathrm{fin}}(T,H = 0\ \mathrm{Oe}) - \kappa_{\mathrm{ini}}(T,H = 0\ \mathrm{Oe})}{\kappa_{\mathrm{ini}}(T,H = 0\ \mathrm{Oe})}, \qquad (2)$$

where $\kappa_{ini}$ and $\kappa_{fin}$ are $\kappa$ before and after experiencing $H > H_c$, respectively. When $H$ exceeds an upper critical field $H_{c1}$ of type-II superconductors, quantum magnetic vortex is formed leading to the decrease of $\kappa_{ph}$ due to the enhanced phonon scattering and increase of $\kappa_{el}$ due to the appearance of normal conducting regions. Thus, the nonvolatile behavior of magneto-thermal conductivity in type-II superconductors reflects the pinning behavior of quantum magnetic vortex. Interestingly, by compositing two superconductors with different $H_c$, even type-I superconductors exhibit an obvious nonvolatile feature. When applied $H$ once exceeds $H_c$ and decreases, magnetic flux is trapped in the regions with smaller $H_c$ preserving $\kappa_{el}$ even at $H = 0$ Oe. A Pb-Sn solder was reported to exhibit the highest nonvolatile MTSR up to 300% owing to the large amount of trapped magnetic flux in Sb region which also breaks the superconducting state in Pb region.[16] Nevertheless, this record value is still one order of magnitude smaller than conventional MTSR in metallic superconductors, which is a barrier toward the application of nonvolatile MTS for energy-efficient thermal management.

We then propose a concept to introduce strong magnetic pinning centers in superconductors as a promising route for giant nonvolatile MTS. From 1960s, an idea to enhance pinning strength via direct magnetic dipole interaction between vortex and ferromagnetic particles has been intensively explored aiming for the enhancement of critical current density.[20–27] Importantly, the strong pinning state in these systems is stabilized when the magnetization of ferromagnets and magnetic field of vortex are oriented in the same direction according to the London approximation,[23] resulting in the enhancement of remanent magnetization.[24] Thus, one naturally expects giant nonvolatile MTSR in superconductor-ferromagnet hybrids as illustrated in **Figure 1**, where ferromagnetic particles are dispersed in superconducting matrix. $\kappa$ initially shows a low value at $T < T_c$ and $H = 0$ due to superconductivity in the matrix (Figure 1a) and increases when $H$ exceeds $H_c$ of the matrix (Figure 1b). After the removal of $H$, the strong magnetic pinning centers is expected to be formed and cause a destruction of larger part of superconducting state than those for the previous type-II and phase-separated superconductors (Figure 1c).

Here, we report the observation of giant nonvolatile MTSR in hybrid systems of superconducting Pb and ferromagnetic Fe. We find that nonvolatile MTS is induced by adding a small amount of Fe (< 1 vol%) and nonvolatile MTSR for the Pb-Fe hybrids monotonically increases with increasing the Fe ratio, which reaches a record-high value of 719% at the Fe ratio of 8.7 vol%, more than twice the value observed in the Pb-Sn solders. The remanent magnetization for the Pb-Fe hybrids is also much larger than that for the Pb-Sn solders, being a significant factor of giant nonvolatile MTS. This work will broaden the framework of materials research and serve as a guideline for hybrid material design for nonvolatile MTS.

## 2. Results and Discussion

### 2.1. Sample Systems

To homogeneously disperse Fe particles in Pb matrix, we use an accumulative roll bonding (ARB) method.[28–33] Pure Pb is known as a type-I superconductor exhibiting large MTSR owing to high $\kappa_{el}$ with longer mean free path of electrons, but MTS arises only in the presence of $H$.[34] Pure Fe has $\kappa$ higher than Pb (100–1000 W $K^{-1}$ $m^{-1}$ below 10 K for 5N purity Fe)[35,36] and large spontaneous magnetization (2.15 T)[37] suitable for magnetic pinning centers. According to the phase diagram, Pb and Fe do not form any intermetallic compounds. **Figure 2** shows a schematic of the ARB process to create the Pb-Fe hybrids. To prevent oxidation of Fe particles, it was sandwiched between Pb plates by a pressing machine under an Ar atmosphere [(i) in Figure 2]. Immediately after annealing to soften Pb (ii), the sandwiched sample was put in a hot rolling machine (Oono-roll Corporation) in the air (iii). The rolled sample was cut into two pieces with the same size (iv), stacked (v), annealed (ii) and then rolled again (iii). We repeated this cycle 15 times to homogeneously disperse the Fe particles inside the Pb plates without melting or sintering processes, which may modify the purity of raw materials. Series samples with different Fe ratios were prepared by tuning the amount of embedded Fe powder. The Fe ratio was estimated from the saturation magnetization measured at 300K for the Pb-Fe hybrids, which is purely derived from the Fe particles.

**Figure 3** shows the elemental maps of Pb and Fe in the Pb-Fe hybrids with various Fe ratios observed by a scanning electron microscopy with an energy dispersive X-ray spectroscopy (SEM-EDX). The Fe particles are homogeneously dispersed in the Pb matrix by ARB with 15 cycles and sustains a particle form with the typical size less than 5 μm. As the embedded Fe ratio increases, the area occupied by Fe consistently increases without forming a cluster of particles. We also confirmed that no peaks originating from the other elements in the SEM-EDX spectrum and no significant voids at the interface between the Pb and Fe were observed. Thus, we successfully obtained series samples of the Pb-Fe hybrids as we originally designed in Figure 1.

### 2.2. Magneto-Thermal Conductivity Measurement

Firstly, we measured the $H$ dependence of $\kappa$ both for increasing and decreasing $H$ in the plain Pb as shown in **Figure 4**a. After zero field cooling (ZFC) to $T = 2.0$ K, $\kappa_{ini}$ is approximately 5 W K$^{-1}$ m$^{-1}$ below $H_c$ of ~800 Oe and drastically increases to 12 W K$^{-1}$ m$^{-1}$ due to the transition into the normal conducting state and appearance of $\kappa_{el}$. These $\kappa$ values below and above $H_c$ are lower than those for 5N-Pb but consistent with those for 3N-Pb in the previous reports[15,34] probably due to the decreased mean free path of electrons and phonons by the ARB process which increases the number of interfaces and induces internal strains. A difference between $\kappa_{ini}$ and $\kappa_{fin}$ is not observed although a slight hysteresis appears only around $H_c$ due to the intermediate-mixed state.[11,12,18,19]

Next, we show the $H$ dependence of $\kappa$ for the Pb-Fe hybrids in Figures 4b–f. As well as Figure 4a, $\kappa$ drastically increases around $H_c$ of Pb for increasing $H$, suggesting the superconducting-to-normal conducting transition of Pb. We find that $\kappa$ does not return to $\kappa_{ini}$ below $H_c$ for decreasing $H$ but sustains the higher value, which occurs even by the addition of Fe particles less than 1 vol%. Let us see the Fe ratio dependence of $\kappa_{ini}$, $\kappa$ above $H_c$, and $\kappa_{fin}$. As the Fe ratio increases, $\kappa_{ini}$ which mainly reflects $\kappa_{ph}$ decreases due to the enhanced phonon scattering by the dispersed Fe particles. On the other hand, $\kappa$ above $H_c$ tends to increase owing to the higher $\kappa$ of pure Fe[35,36] and the nonvolatility becomes pronounced as the Fe ratio increase, resulting in the higher $\kappa_{fin}$ values at $H = 0$ Oe.

### 2.3. Observation of Giant Nonvolatile Magneto-Thermal Switching

**Figure 5**a plots $\kappa_{ini}$ and $\kappa_{fin}$ for the Pb-Fe hybrids and Pb-Sn solders[16] as a function of the inclusion (Fe and Sn) ratio. In the Pb-Sn hybrids, as the Sn ratio decreases, the difference in $\kappa_{ini}$ and $\kappa_{fin}$ obviously emerges and $\kappa_{ini}$ decreases owing to lower $\kappa$ of Pb than Sn[14,15] and phonon scattering in the phase-separated composite structure. However, $\kappa_{fin}$ also decreases associated with the suppression of $\kappa_{el}$ as well as $\kappa_{ph}$, being the trade-off relationship to obtain large nonvolatile MTSR. Then, let us see the result for the Pb-Fe hybrids. The $\kappa_{ini}$ values decreased by dispersing Fe particles are comparable to those of the Pb-Sn solders, ensuring the contribution of micro-scale inclusions to $\kappa_{ph}$. According to the results for both systems, $\kappa_{ini}$ minimizes when the inclusion ratio ranges in 10–20 vol%. Contrary to the result for the Pb-Sn solders, $\kappa_{fin}$ drastically increases within this inclusion ratio range owing to the higher $\kappa$ of pure Fe, breaking the trade-off between the enhancement of $\kappa_{el}$ and reduction of $\kappa_{ph}$.

**Figure 5**b plots nonvolatile MTSR for the Pb-Fe hybrids as a function of the inclusion ratio compared with the Pb-Sn solders. The simultaneous achievement of high $\kappa_{fin}$ and low $\kappa_{ini}$ in the Pb-Fe hybrids leads to giant nonvolatile MTSR up to 719% at the Fe ratio of 8.7 vol%. This value is more than twice the previous record nonvolatile MTSR for the Pb-Sn solders with the comparable inclusion ratio and even the volatile MTSR for the 3N purity Pb. [34]

### 2.4. Characterization of Remanent Magnetization

We measured the magnetic property for the Pb-Fe hybrids to characterize the remanent magnetization ($4\pi M_r$) in relation to the observed giant nonvolatility in magneto-thermal conductivity. **Figures 6**a and b show the $T$ dependences of the magnetization ($4\pi M$) from 2 K to 10 K at $H$ = 10 Oe measured immediately after ZFC and field cooling under 1500 Oe (FC) processes, respectively. In the normal conducting state at 10 K, the $H$ dependence of $4\pi M$ derived from the Fe particles has no hysteresis behavior. In Figure 6a, all the samples exhibit $4\pi M$ = -10 G below $T_c$ of Pb (7.2 K) due to the perfect diamagnetism and positive values above $T_c$ due to the ferromagnetism of Fe, which monotonically increase as the Fe ratio increases. On the other hand, after FC, $4\pi M$ exhibits positive values even below $T_c$ (Figure 6b), which are two orders of magnitude larger than those solely by ferromagnetism of Fe particles.

Figure 6c plots $4\pi M_r$, which is the $4\pi M$ value at $T$ = 2.0 K in Figure 6b, for the Pb-Fe hybrids as a function of the inclusion ratio in comparison with that for the Pb-Sn solders. $4\pi M_r$ for the Pb-Fe hybrids monotonically increases as the Fe ratio increases and is much larger than that of the Pb-Sn solders at the comparable inclusion ratio. In the Pb-Sn solders, $4\pi M_r$ is due to the trapped magnetic flux in the Sn region with the size and proximity effects.[38] However, the superconducting proximity effect is hardly induced in the Fe region because the magnetic exchange field strongly opposes the formation of Cooper pairs. Thus, the large $4\pi M_r$ for the Pb-Fe hybrids cannot be explained solely by the trapped magnetic flux or ferromagnetic moment in the Fe region. A possible mechanism is that, when $H$ decreases below $H_c$ at $T$ = 2.0 K, the vortex clusters could be formed in the Pb region at the boundary with the Fe particles due to the spatially inhomogeneous magnetization.[27] The vortex clusters, in turn, induce the magnetic field localized at the boundary and hence finite ferromagnetic moment in Fe even at zero external magnetic field. Thus, the vortex clusters and ferromagnetic moment can concertedly enhance the total $4\pi M$ and form the strong magnetic pinning centers due to the dipole interaction, resulting in the anomalously large $4\pi M_r$.

Finally, let us discuss the relationship between $4\pi M_r$ and the nonvolatile MTS. To refer to how high the $\kappa_{fin}$ value sustains close to $\kappa$ above $H_c$, we introduce a nonvolatility factor as,

$$\text{Nonvolatility factor} = \frac{\kappa_{\text{fin}}(T, H = 0\ \text{Oe}) - \kappa_{\text{ini}}(T, H = 0\ \text{Oe})}{\kappa(T, H = 1500\ \text{Oe}) - \kappa_{\text{ini}}(T, H = 0\ \text{Oe})}. \tag{3}$$

Figure 6d plots the nonvolatility factor as a function of $4\pi M_r$ for the Pb-Fe hybrids and Pb-Sn solders.[16] Both the systems show a similar trend that the nonvolatility factor linearly increases with increasing $4\pi M_r$ and saturates around 0.7–0.8 above a threshold value ($4\pi M_r \sim 700$ Oe).

Thus, the enhancement of $4\pi M_r$ is indispensable to the emergence of the large nonvolatility in the magneto-thermal conductivity in superconducting composite systems. Note that the achievement of nonvolatility factor of 1 is impossible in superconductor-ferromagnet hybrid systems because a superconducting region with low $\kappa$ is required to form and strongly trap magnetic flux inside the systems.

**3. Conclusion**

In conclusion, we demonstrated giant nonvolatile MTS at low temperatures in superconductor-ferromagnet hybrids. The ferromagnetic Fe particles were homogeneously dispersed in superconducting Pb matrix by means of the ARB method. The magnetic field dependence of $\kappa$ revealed that nonvolatile MTS was induced by dispersing a small amount of Fe particles and nonvolatile MTSR reached a record-high value of 719% at Fe ratio of 8.7 vol%. The importance of our concept lies on that the dispersed Fe particles simultaneously increase $\kappa_{el}$ and decrease $\kappa_{ph}$ and, in concert with the trapped magnetic flux, the ferromagnetic moment of Fe breaks the superconductivity of the Pb matrix. This work will broaden the materials research for nonvolatile MTS enabling energy-efficient control of heat transfer.

**4. Methods**

*Sample preparation*: Pb plates with 5N purity (The Nilaco Corp.) and Fe powder with 3N purity (Kojundo Chemical Lab. Co. Ltd.) were used as raw materials. The size of Fe particle was selected as 3–5 μm. The Pb plates were originally cut into pieces with a dimension of 15 mm × 20 mm × 0.5 mm. At the ARB process (ii), the samples were annealed for 2 min at 220°C in an electric furnace under a nitrogen atmosphere. At the ARB process (iii), the distance between upper and lower rollers was set to be 0.2 mm with the surface temperature of 150°C and rolling speed of 1.0 m/min. A typical size of the Pb-Fe hybrids after the ARB process was 105 mm × 20 mm × 0.2 mm. For the measurements of $\kappa$ and magnetization, the samples were cut into dimensions of 25 mm × 1.0 mm × 0.2 mm and 5.0 mm × 1.0 mm × 0.2 mm, respectively.

*Measurement methods*: A Cross-Beam 1540ESB (Carl Zeiss AG) was used for the SEM-EDX observations. The cross-section surface of the samples was mechanically polished in advance. A thermal transport option of Physical Property Measurement System (PPMS, Quantum Design) was used to measure $\kappa$ by a steady-state method. All the $\kappa$ measurements were performed after ZFC from ≧10 K and stabilization at 2 K in the PPMS chamber at high vacuum state and while sweeping $H$ from 0 to 1500 Oe and from 1500 to 0 Oe. Magnetic

Property Measurement System (MPMS, Quantum Design) was used for all the magnetization measurements.

**Acknowledgements**

The authors thank H. Sebata, M. Maeda, and N. Kurata for technical supports. This work is supported by JST-ERATO “Magnetic Thermal Management Materials” (Grant No. JPMJER2201), Japan, and Iketani Science and Technology Foundation.

**Author Contributions**

F.A. and K.U. planned and supervised the study. Y.S. prepared the samples and performed all the measurements. K.I., Y.W., P.R. and Y.M. supported the thermal conductivity measurement and H.S.A. supported the microstructure analysis. Y.S. and F.A. developed the explanation of the experimental results with the help from T.H. All the authors discussed the results and commented on the manuscript.

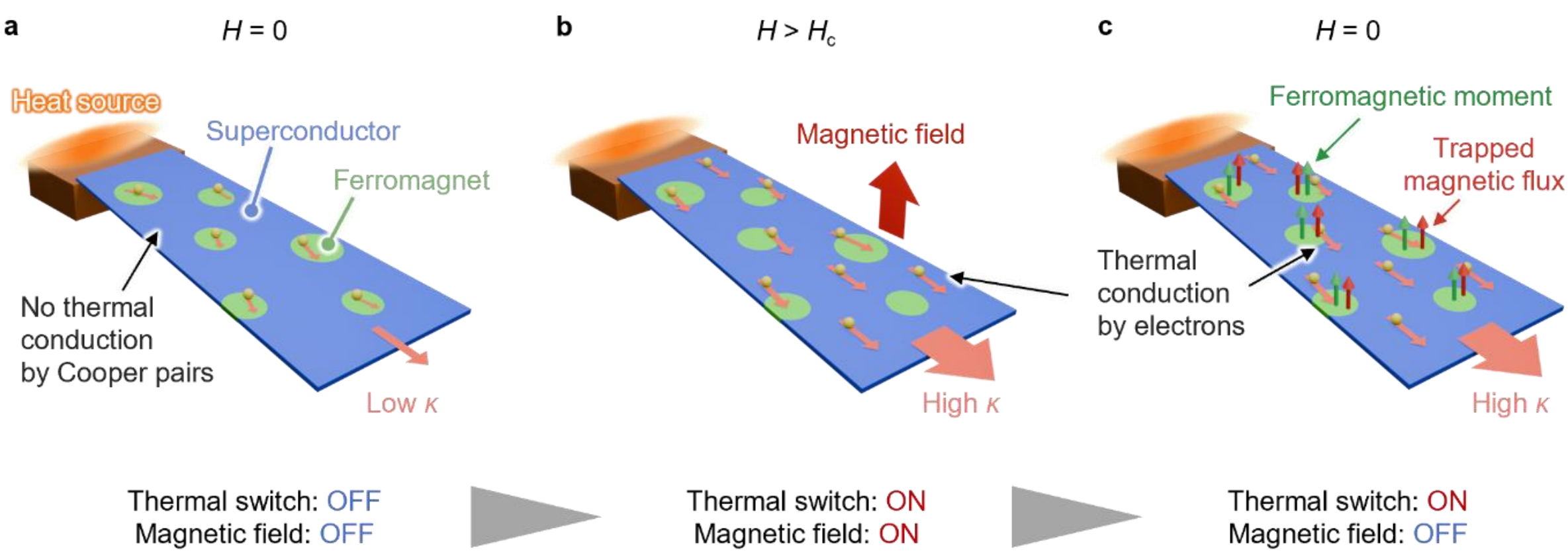

**Figure 1.** Schematic of nonvolatile magneto-thermal switching (MTS) in a superconductor-ferromagnet hybrid, where ferromagnetic particles are dispersed in a superconductor matrix. a) Thermal conductivity $\kappa$ exhibits a low value at an external magnetic field $H$ = 0 due to superconductivity of the matrix. b) $\kappa$ increases when $H$ exceeds the critical field $H_c$ of the matrix due to the appearance of electron contribution. c) Even after the removal of $H$, ferromagnetic moment in concert with the trapped magnetic flux is expected to break larger part of superconducting state, preserving the higher $\kappa$ value.

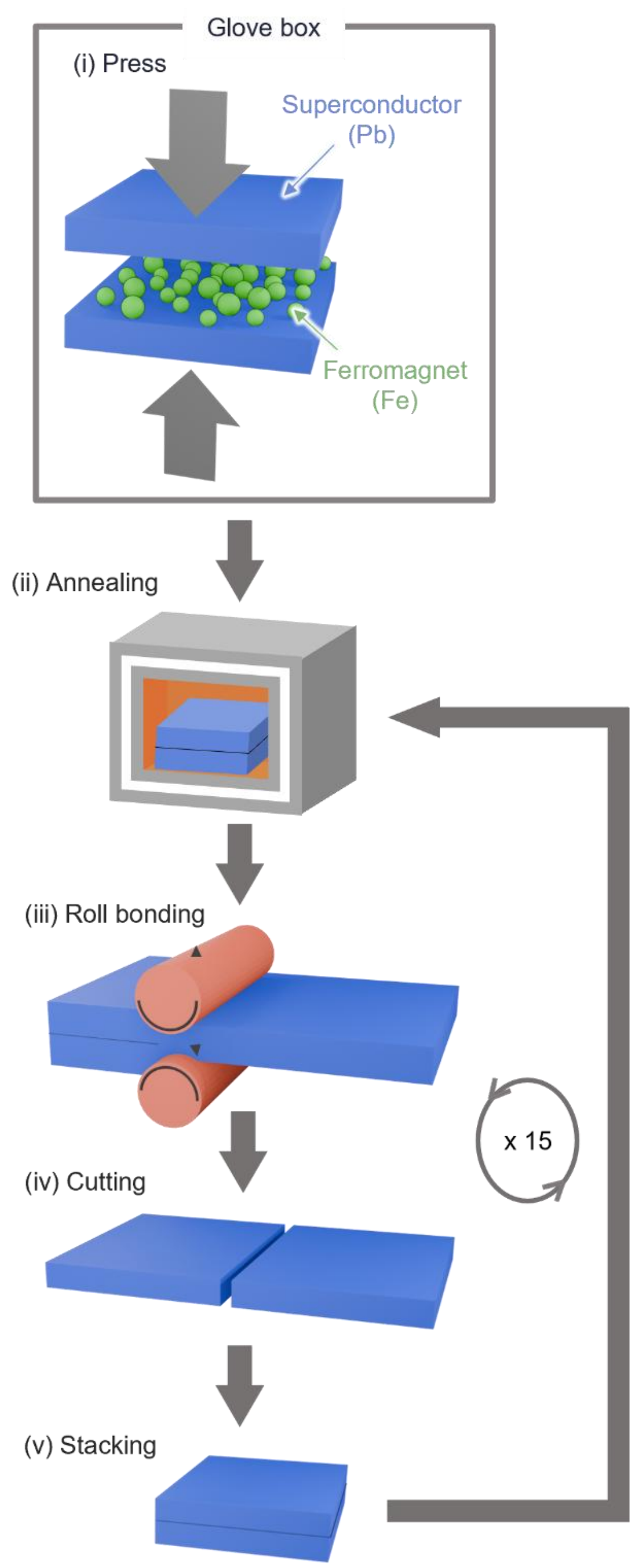

**Figure 2.** Schematic of an accumulative roll bonding method to create the Pb-Fe hybrids. (i) Fe particles are sandwiched between Pb plates by a pressing machine under an Ar atmosphere. To homogeneously disperse the Fe particles inside the Pb plates, we repeated a cycle of (ii) annealing, (iii) hot rolling, (iv) cutting into two pieces, and (v) stacking for 15 times.

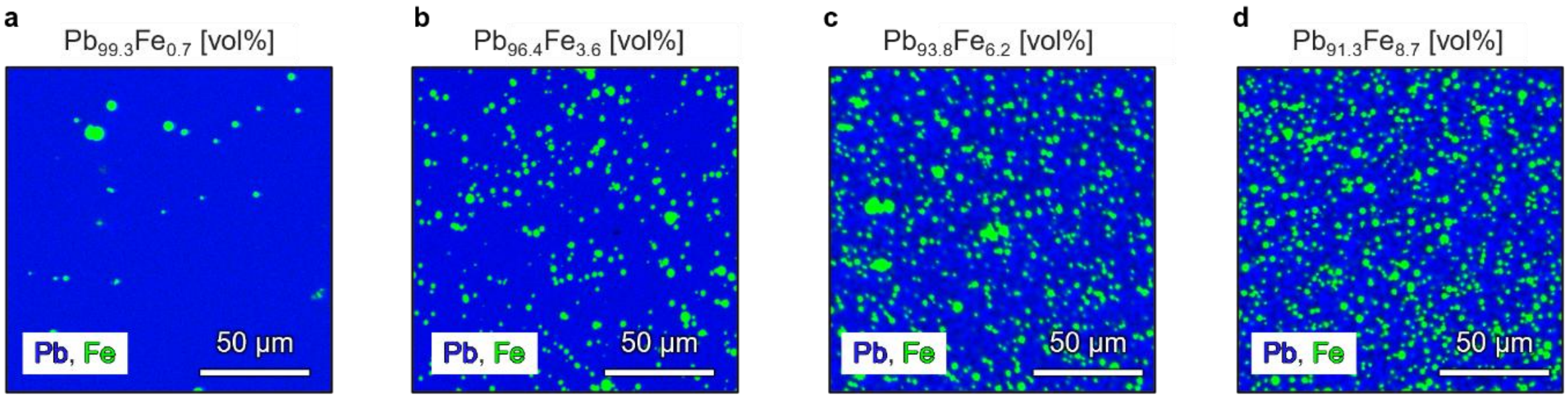

**Figure 3.** Elemental maps of Pb and Fe in the Pb-Fe hybrids with the Fe ratio of a) 0.7, b) 3.6, c) 6.2, and d) 8.7 vol% observed by a scanning electron microscopy with an energy dispersive X-ray spectroscopy.

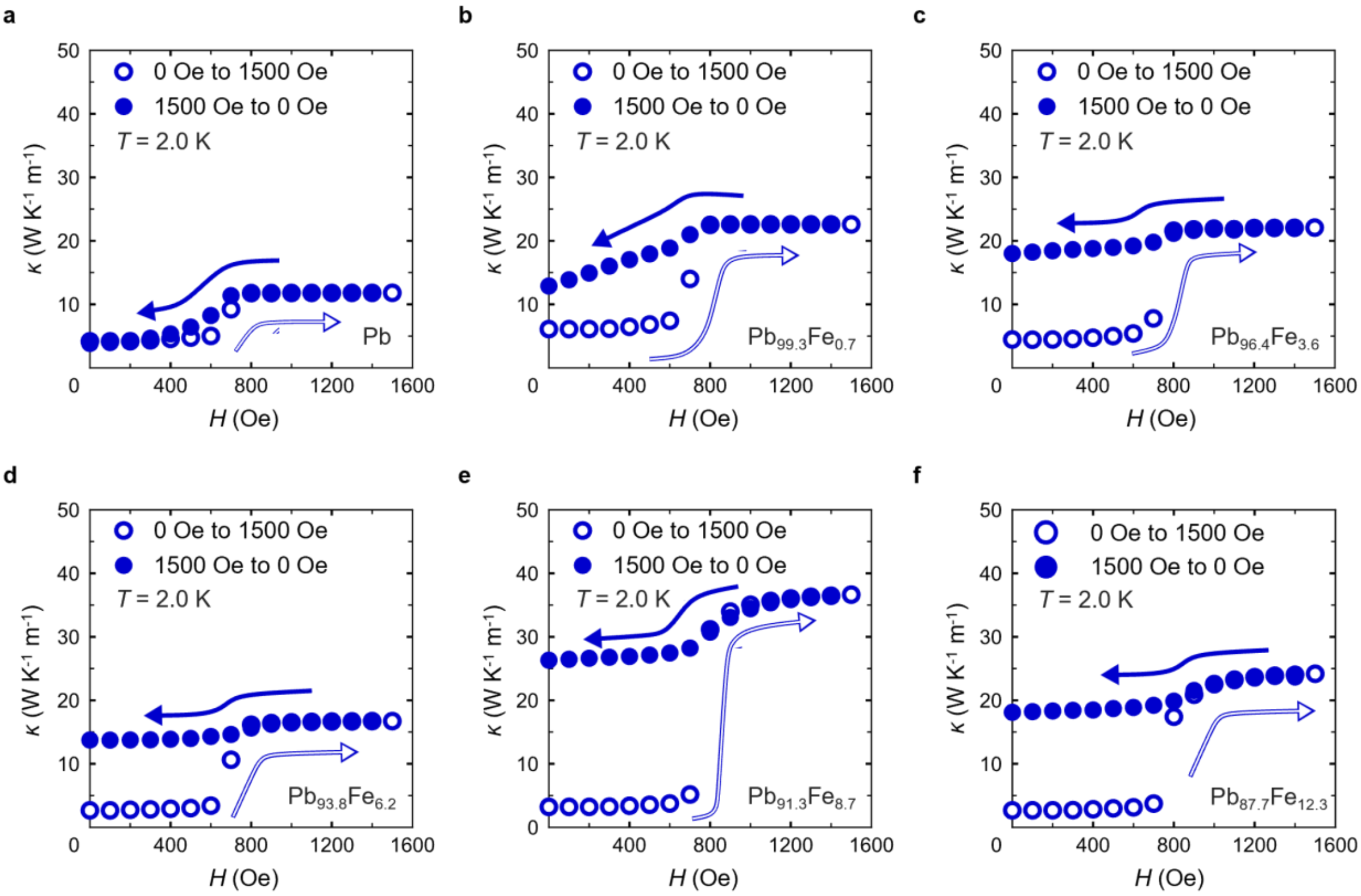

**Figure 4.** The $H$ dependence of $\kappa$ at the temperature $T$ = 2.0 K both for increasing and decreasing $H$ in a) the plain Pb and b–f) Pb-Fe hybrids with the Fe ratio of b) 0.7, c) 3.6, d) 6.2, e) 8.7, and f) 12.3 vol%.

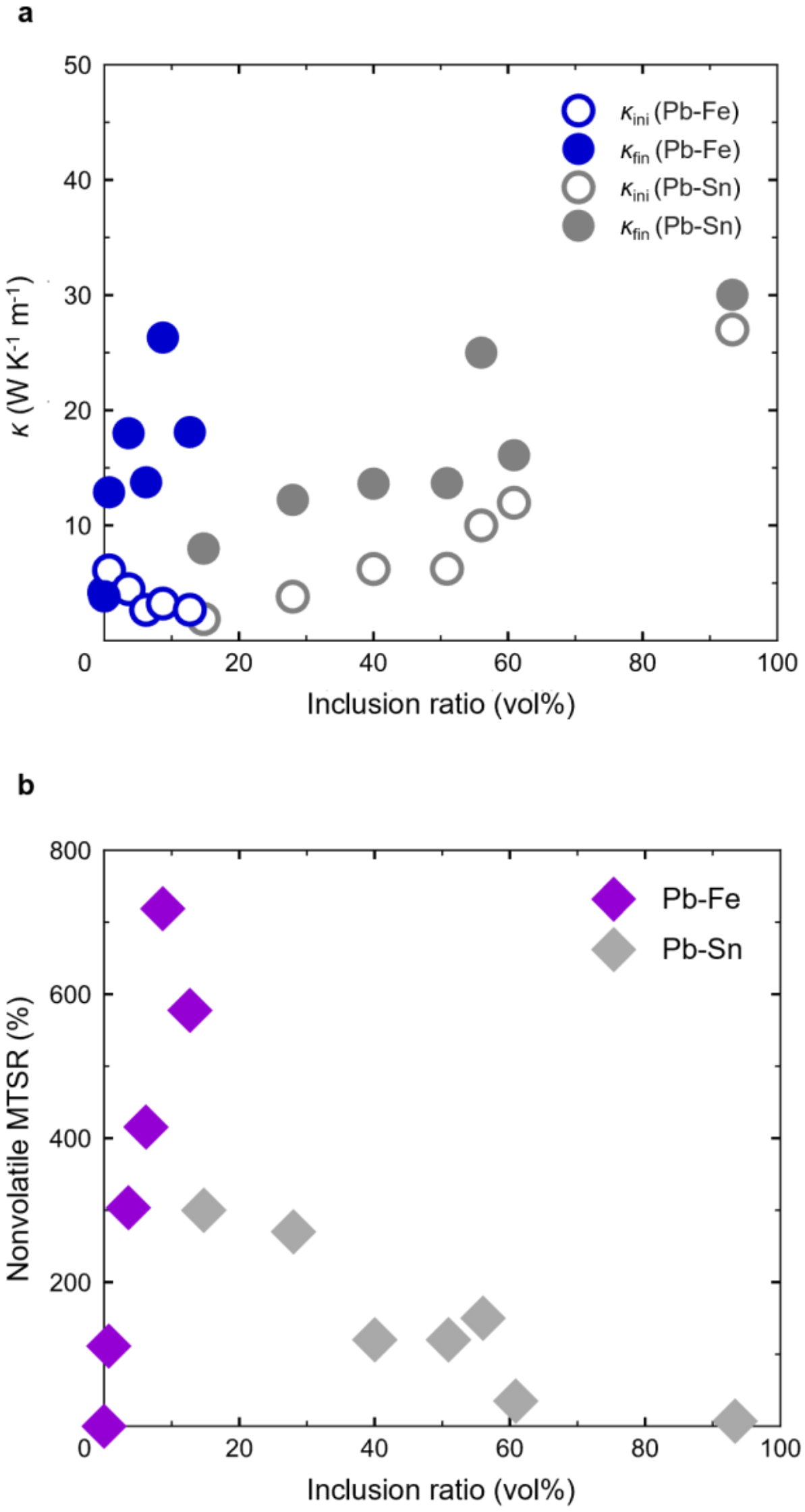

**Figure 5.** Observation of giant nonvolatile MTS. a) $\kappa_{ini}$ and $\kappa_{fin}$ and b) nonvolatile magneto-thermal switching ratio (MTSR) as a function of the inclusion (Fe and Sn) ratio for the Pb-Fe hybrids and Pb-Sn solders[16].

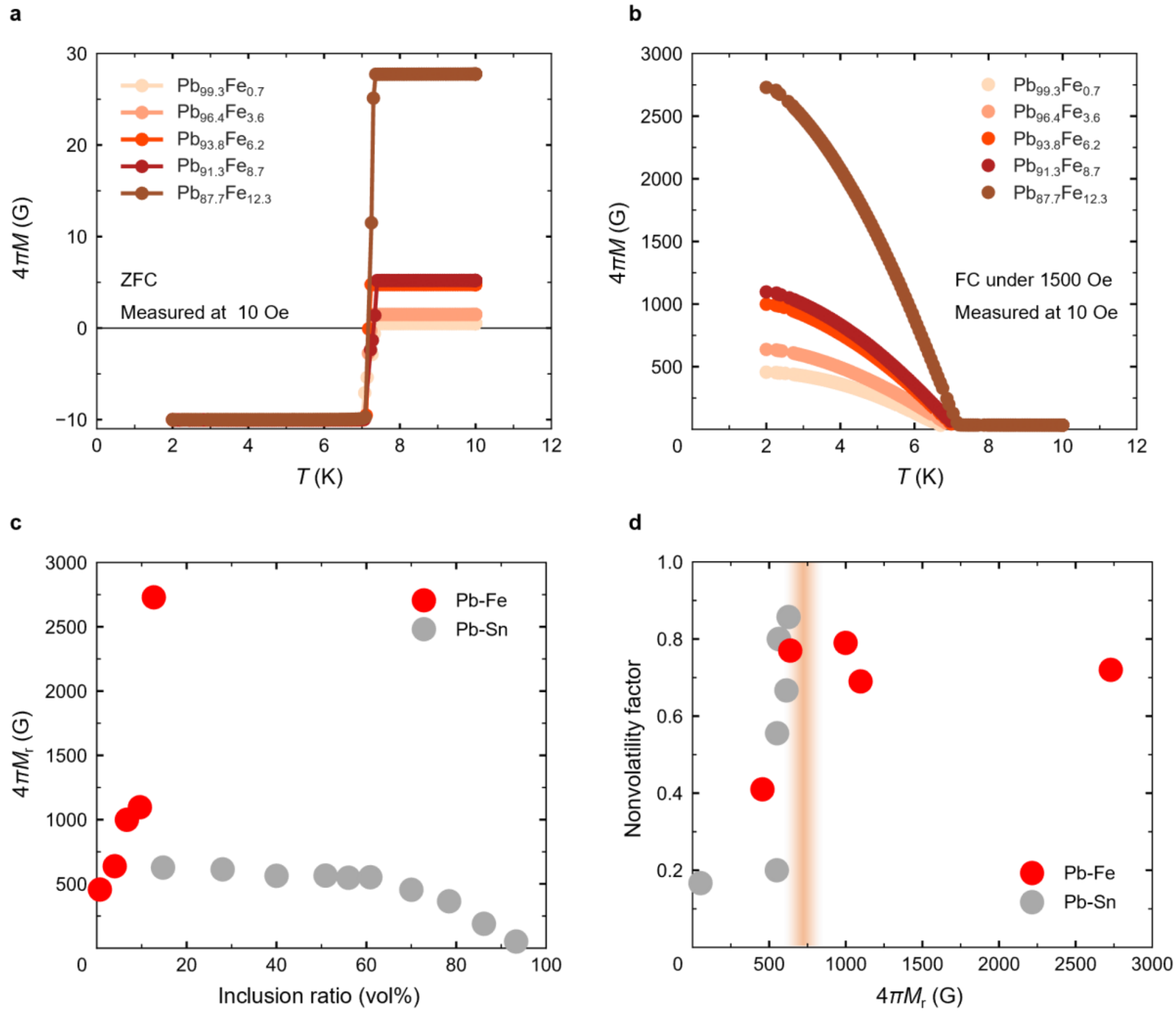

**Figure 6.** a,b) The $T$ dependence of the magnetization $4\pi M$ at $H = 10$ Oe measured immediately after a) zero field cooling (ZFC) and b) field cooling under 1500 Oe (FC) processes. c) The remanent magnetization $4\pi M_r$ as a function of the inclusion (Fe and Sn) ratio and d) the relationship between the nonvolatility factor and $4\pi M_r$ for the Pb-Fe hybrids and Pb-Sn solders[16].